\documentclass[aps,pra,twocolumn,eqsecnum,superscriptaddress]{revtex4}

\usepackage{epsfig}

\begin{document}

\title{Quantum information science as an approach to complex quantum systems}

\author{Michael~A.~Nielsen}
\email[Email: ]{nielsen@physics.uq.edu.au}
\homepage[\\ Homepage: ]{http://www.qinfo.org/people/nielsen/}

\affiliation{Department of Physics, University of Queensland,
Queensland 4072, Australia}
\date{\today}

\begin{abstract}
  What makes quantum information science a \emph{science}?  These
  notes explore the idea that quantum information science may offer a
  powerful approach to the study of \emph{complex quantum systems}.
  We discuss how to quantify complexity in quantum systems, and argue
  that there are two qualitatively different types of complex quantum
  system.  We also explore ways of understanding complex quantum
  dynamics by quantifying the \emph{strength} of a quantum dynamical
  operation as a physical resource.  This is the text for a talk at
  the ``Sixth International Conference on Quantum Communication,
  Measurement and Computing'', held at MIT, July 2002.  Viewgraphs for
  the talk may be found at ``http://www.qinfo.org/talks/''.
\end{abstract}

\maketitle

\section{Introduction}

Good afternoon, ladies and gentlemen.  My name is Michael Nielsen and
I'm from the University of Queensland.  My talk today is about using
quantum information science as an approach to the study of complex
quantum systems.  The work I will describe has involved many
collaborators at the University of Queensland and MIT, but I would
especially like to emphasize the key role played by Tobias Osborne in
this research.

Let me begin by asking what it is that makes quantum information
science a science?  Friends outside the field sometimes comment that
it seems to be largely engineering, with little science.  A standard
response to this question from physicists has been that in the course
of building devices like quantum computers, we'll discover lots of
interesting and fundamental physics.

This is undoubtedly true, and is an excellent reason for doing quantum
information science.  But it seems a little like the argument
sometimes used to justify going to the moon, namely, that it resulted
in valuable spin-off technologies in fields such as computation and
aeronautical engineering.  Of course, this misses a large part of the
point, since going to the moon has an \emph{intrinsic worth}, a point
obvious to those with even the slightest romance in their soul.

What is the intrinsic worth of quantum information science?  In my
talk today I'll argue that quantum information science may be regarded
as an approach to the study of \emph{complex quantum systems}.
Related ideas have been advocated by many people.
Aharonov~\cite{Aharonov99b}, Nielsen~\cite{Nielsen98d} and
Preskill~\cite{Preskill00a} argued that there may be interesting
connections between the quantitative theory of entanglement and the
properties of many-body quantum systems, and there is now a burgeoning
literature exploring these connections~\footnote{See, for
  example~\cite{Osborne02a,Osterloh02a,Wootters02a,Gunlycke01a,Wang02a}
  for a selection of recent work and further references.}.

More explicitly, the concluding paragraph of
DiVincenzo's~\cite{Divincenzo00a} paper on the physical requirements
for quantum computation suggests that quantum information science may
offer valuable insights into complex quantum systems.  This theme was
explored in more detail by Osborne and
Nielsen~\cite{Nielsen01d,Osborne02b,Osborne02a,Osborne02c}, and the
present talk is an outgrowth of this work, with some illustrative
examples drawn from~\cite{Bremner02a,Nielsen02a}.

\section{Complex quantum systems}

What is a complex system?  Complexity is an elusive concept: it is
difficult to define, but we know it when we see it.  In response to
this difficulty one is naturally led to ask whether it is possible to
come up with quantitative measures of complexity.

In the 1980s Bennett proposed just such a measure, which he called the
\emph{logical depth} of a system~\cite{Bennett88b,Bennett90a}.  The
basic idea is that a system should be called complex, or logically
deep, if that system can be generated by a few simple rules, but those
rules require a long time to run.  So, for example, a human body is
complex in that it is specified by a relatively small amount of
information encoded in DNA, but it takes a great deal of processing to
get from that DNA to the human body.

Another example is a regular pattern on a checkerboard, which is not
complex because it can be quickly generated by a simple rule.  More
subtle is the case of a random pattern on a checkerboard.  That is not
complex either, because there is no simple rule generating the
pattern.  Indeed, the simplest rule generating the pattern is simply
the program which contains (and prints) a complete listing of the
states of all the elements of the checkerboard, and this program runs
very quickly.

Now let me give you an example of something complex, that is, with
high logical depth.  Suppose I take a point $x$ in the plane, for
example, $x=(0,0)$.  I now bounce the point around the plane by
applying one of the following four rules at random, over and over
again~\cite{Barnsley88a},
\begin{eqnarray}
& x \rightarrow \left[ \begin{array}{cc} 
    0 & 0 \\ 0 & 0.16 \end{array} \right]x &
\mbox{with probability } 0.01   \nonumber \\
 & & \label{eq:fern1} \\
 & x \rightarrow \left[ \begin{array}{cc}  
    0.85 & 0.04 \\-0.04 & 0.85 \end{array} \right]x 
    + \left[ \begin{array}{c} 0 \\ 1.6 \end{array} \right]
    & \mbox{with probability } 0.85 \nonumber \\ & & \\
 & x \rightarrow \left[ \begin{array}{cc} 
    0.2 & -0.26 \\ 0.23 & 0.22 \end{array} \right]x 
    + \left[ \begin{array}{c} 0 \\ 1.6 \end{array} \right]
    & \mbox{with probability } 0.07 \nonumber \\ & & \\
 & x \rightarrow \left[ \begin{array}{cc} 
    -0.15 & 0.28 \\ 0.26 & 0.24 \end{array} \right]x 
    + \left[ \begin{array}{c} 0 \\ 0.44 \end{array} \right]
    & \mbox{with probability } 0.07. \nonumber \\
 & & \label{eq:fern4}
\end{eqnarray}

\begin{figure}[t]
\epsfig{file=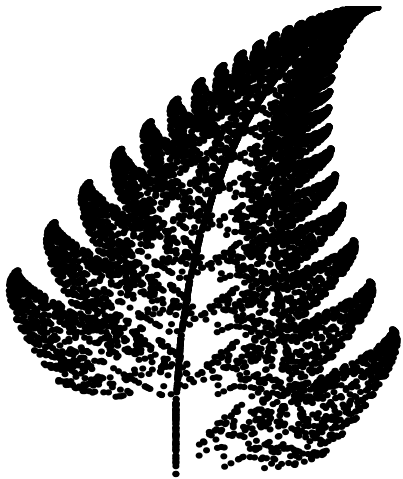}
\caption{The fern shape is formed by iterating the simple rule 
in Eqs.~(\ref{eq:fern1})-(\ref{eq:fern4}) a few thousand times.
\label{fig:fern}}
\end{figure}

When you repeat this simple procedure a few thousand times a very
interesting thing starts to happen: a fern shape starts to fill in, as
illustrated in Fig.~\ref{fig:fern}.  You can prove that this happens
with very high probability: the fern is an ``attractor'' for this
stochastic dynamical system.  What is more, the fern is complex in
that there is a simple rule generating the fern, but it takes a long
time to run.

Bennett formalized these intuitions by defining the logical depth of a
data set to be the running time of a near-optimal computer program
generating that data set.  There are some technicalities hidden in
this definition, like the precise meaning of ``near-optimal'', that
I'm going to gloss over today.  Nonetheless, I would like to give you
an intuitive feel for what the definition means.

First of all, by ``near-optimal'' we mean that the program is nearly
the shortest possible.  The motivating idea for this requirement is an
analogy between computer programs generating data sets and scientific
hypotheses.  Scientists tend to prefer simple explanations over more
complex; this is Occam's razor.  If we think of computer programs as
explanations for data sets, then we would tend to prefer short
computer programs --- simple ``explanations'' --- over longer
programs.

With this definition, simple repeated structures and random patterns,
like the checkerboards I described earlier, have low logical depth.
However, systems like the fern have high logical depth because they
have very simple explanations that take a long time to run.

Up until now I've said nothing about quantum mechanics.  However, it
turns out that there is an interesting quantum twist to logical depth.
As we know, factoring integers is a hard problem, hard enough that RSA
systems offers lots of money to people able to factor integers like
this collossus\footnote{RSA Systems offers US \$200,000 for the
  following number, at the time of this writing.}:
\begin{eqnarray}
& & 25195908475657893494027183240048398571429282 \nonumber \\
& & 12620403202777713783604366202070759555626401 \nonumber \\
& & 85258807844069182906412495150821892985591491 \nonumber \\
& & 76184502808489120072844992687392807287776735 \nonumber \\
& & 97141834727026189637501497182469116507761337 \nonumber \\
& & 98590957000973304597488084284017974291006424 \nonumber \\
& & 58691817195118746121515172654632282216869987 \nonumber \\
& & 54918242243363725908514186546204357679842338 \nonumber \\
& & 71847744479207399342365848238242811981638150 \nonumber \\
& & 10674810451660377306056201619676256133844143 \nonumber \\
& & 60383390441495263443219011465754445417842402 \nonumber \\
& & 09246165157233507787077498171257724679629263 \nonumber \\
& & 86356373289912154831438167899885040445364023 \nonumber \\
& & 52738195137863656439121201039712282212072035 \nonumber \\
& & 7.
\end{eqnarray}

Let's optimistically imagine, then, that it's ten years from now and
somebody wants to prove that they have a functioning quantum computer
in their lab, but don't want to reveal all the details of how they
built it.  One good way of proving this to the world would be to
publish a paper~\cite{Nobelist12a} containing the prime factors of a
large group of big integers --- perhaps the prime factors of all
numbers between $10^{500}$ and $10^{500}+100000$.

Now, would this list of factors constitute a complex system or not?
The answer is that it depends on whether the computer in the
definition of logical depth is quantum or classical.  If it's quantum
then it seems likely that this system is \emph{not} logically deep,
and thus is not complex, because we can quickly generate the list
using a short quantum program, namely, Shor's
algorithm~\cite{Shor94a,Shor97a}.  If the definition of logical depth
uses a classical computer, and there really is, as we suspect, no fast
classical factoring algorithm, then the list of factors has high
logical depth, since there are simple computer programs capable of
generating such a list, but they operate very slowly.

Thus, according to this discussion there are two distinct notions of
logical depth, classical logical depth, and quantum logical depth.  We
can summarize the situation by dividing the class of possible data
sets into three distinct types.  First, there are data sets which have
both low classical logical depth and low quantum logical depth.  Then
there are data sets that have high classical logical depth, but low
quantum logical depth, like the list of prime factors that I
discussed.  Then there are data sets with both high classical logical
depth and high quantum logical depth.  I don't yet have a really good
example of a system which I expect to have this property, but consider
some systems likely candidates, for example, the output of a quantum
cellular automata that's been running for a long time.  Note that
systems with high quantum logical depth but low classical logical
depth seem unlikely, because a simple, fast classical computer
generating a data set can be simulated by a simple, fast quantum
computer.

We thus conclude that there are two different types of complex system,
one which we might call semi-classically complex, being systems with
high classical logical depth, but low quantum logical depth.  The
other we might call quantum complex, having both high quantum and
classical logical depth.

\section{Quantum dynamics as a physical resource}

I've talked a fair bit about quantifying complexity, and the
information-theoretic viewpoint has led us to the idea that there are
at least two, if not more, qualitatively different types of complex
system.  I'd like now to talk about what we can learn about specific
complex quantum systems from quantum information science.

As you are all aware, over the past few years a great deal of effort
has been devoted to developing a quantitative theory of quantum
entanglement~\footnote{See, for example, the recent review articles
  \cite{Horodecki01a,Horodecki01b,Wootters01a,Nielsen01b} for an
  introduction and further references.}.  In my opinion one of the
major areas in which this theory will be applied is to obtain insights
into the static properties of complex quantum systems.  We're already
starting to see this with applications of the theory of entanglement
to condensed matter systems~\footnote{See, for
  example~\cite{Osborne02a,Osterloh02a,Wootters02a,Gunlycke01a,Wang02a}
  for a selection of recent work and further references.}.  Indeed,
when I originally planned this talk I was going to spend about a third
of it discussing how the theory of entanglement provides insight into
an important numerical tool for studying many-body quantum systems,
the so-called density matrix renormalization group~\cite{Osborne02c}.

However, in the meantime my group has obtained what I think are some
more interesting results that can be used to illustrate how quantum
information science provides insight into complex quantum systems, so
I will talk about those instead.

These other results arise from the observation that static quantum
entanglement is only a small part of the story: it would also be
interesting to obtain a better understanding of the quantum
\emph{dynamics} of complex systems.  To do this, we've focused on the
question of whether or not it might be possible to quantify the
\emph{strength} of a quantum dynamical operation for quantum
information processing.

The motivation behind this idea is the observation, made last year,
that quantum dynamics are a \emph{fungible} physical resource.  By
fungible, I mean that it is possible to convert different types of
dynamical operation, one to the other, just as it is possible to
convert one type of entangled state to another.

Let me be a little more precise about what I mean.  Suppose a system
contains $n$ qudits, that is $d$-dimensional quantum systems, where $d$
can be any integer.  We further suppose that the Hamiltonian for the
system contains only two-body terms, and so can be represented by a
graph with the vertices representing qudits, and an edge between
vertices representing the presence of an interaction between those
qudits.  We'll further suppose that the graph is connected, so that
different qudits aren't completely cut off from one another.

What is interesting is that by alternating evolution due to such a
Hamiltonian with single-qudit gates, it turns out that we can
efficiently simulate any quantum computation, at least in
principle~\footnote{Many results in this vein have been obtained over
  the past 18 months or so.  Further information on these results may
  be found
  in~\cite{Dodd02a,Dur01a,Bennett01a,Wocjan02a,Wocjan02b,Nielsen01e,Vidal01b},
  and references therein.}.  For our purposes, the interesting
conclusion to draw from this discussion is that these Hamiltonians
form a fungible physical resource, since it is possible to use any one
to simulate any other.

Now, this is a very interesting theoretical result, however one might
ask whether the result is practically useful for quantum computation.
Certainly the schemes proposed last year were not practically useful,
since they required extremely frequent and rapid local control to do
the simplest of operations, such as a controlled-{\sc not}.  Even an
extremely optimistic example~\cite{Dodd02a} required something like
$10^4$ operations to do a single controlled-{\sc not}.

Recently, however, the situation has changed.  Imagine one is given a
single entangling two-qubit unitary operation, $U$, and is asked to
perform a controlled-{\sc not} using just $U$ and local unitaries.
Last year, J.-L. and R. Brylinski~\cite{Brylinski02a} were able to
show that this is always possible.  Indeed, they were even able to
generalize the result to the qudit case.  However, they needed to use
quite a few results from algebraic geometry and the theory of Lie
algebras to prove their results, and it is not clear to me whether
their proof can be used to give an efficient constructive method for
doing the controlled-{\sc not}.

However, just a couple of weeks ago, my group put a paper on the
archive~\cite{Bremner02a} that provides a \emph{simple and
  constructive algorithm} for doing a controlled-{\sc not} using these
resources.  Perhaps more importantly, the algorithm even turns out to
be \emph{near-optimal}, in the sense that it uses what is very nearly
the minimal number of uses of $U$ to simulate a controlled-{\sc not}.
An online implementation of this algorithm, due to Chris Dawson and
Alexei Gilchrist, may be found at
``http://www.physics.uq.edu.au/gqc/''.  This algorithm shows that not
only are quantum dynamical operations fungible in principle, but they
may be fungible in practice as well.

Indeed, I believe that this algorithm is one step along the way in an
interesting evolution in the viewpoint of a would-be quantum computer
designer.  In the early days of quantum computing, the question for
such a designer was always ``How can I quantum compute, given the
interactions in my system?''  What I hope our result and other similar
results will do is change that question so that it becomes simply
``What is the interaction in my system?'', with the method for doing
quantum computation in an optimal way simply pulled up out of a
database once the interaction is known.

Knowing that quantum dynamics are a fungible physical resource, and
thus qualitatively equivalent to one another, it makes sense to ask
whether we can quantify the \emph{strength} of a particular dynamical
operation.  The picture to have in mind is an interaction, $U$, being
applied to $n$ qubits.  We then attempt to quantify the strength by
some appropriate function, $K(U)$.  The letter $K$, by the way, comes
from the Greek God of strength and power, Krakos.  (Note that the
following ideas about quantifying dynamic strength are based on a
paper by Nielsen, Dawson, Dodd, Gilchrist, Mortimer, Osborne, Bremner,
Harrow, and Hines~\cite{Nielsen02a}, and much more discussion can be
found there.)

Let me give you a couple of examples of strength measures.  These are
just two examples chosen on an \emph{ad hoc} basis from the much
larger collection of strength measures considered
in~\cite{Nielsen02a}.  For simplicity I will state both measures just
for two qubits, although extensions to more than two qubits are easily
defined.  The first measure of strength, the entangling power of a
unitary operation, is just the maximal change in entanglement that the
unitary can cause, maximized over all possible pure input states,
\begin{eqnarray}
K_{\Delta}(U) \equiv \max_{\psi} |E(U\psi)-E(\psi)|,
\end{eqnarray}
where $E(\cdot)$ is some appropriate measure of entanglement.  The
second measure of strength is very different in nature.  Imagine we
have a metric $D$ on the space of all unitary operations.  Then this
metric induces a natural strength measure, $K_D$, as follows:
\begin{eqnarray}
K_D(U) \equiv \min_{A,B} D(U,A\otimes B).
\end{eqnarray}
That is, the strength is just the minimal distance between $U$ and the
set of local unitaries.  (Note that similar approaches to the
definition of entanglement measures have been explored
in~\cite{Vedral97a,Vedral98a}.)

What good are these measures of strength, even assuming we could
obtain useful computational formulas for them?  Let me answer that
question by talking about an interesting connection between measures
of dynamic strength and computational complexity.  Computational
complexity theory is concerned with the problem of determining how
many elementary quantum gates are needed to perform a particular
unitary $U$.  In other words, it's about quantifying the complexity,
not of data sets, as we talked about earlier, but rather the
complexity of particular dynamical operations.

It turns out that there's an interesting relationship between strength
measures and computational complexity.  Imagine we had a strength
measure with the following three properties.

The first property, which I will call chaining, just says that the
strength of a product of two unitary operations, $U$ and $V$, must be
less than the sum of their combined strengths.  That is,
\begin{eqnarray}
K(UV) \leq K(U)+K(V).
\end{eqnarray}
The intuition behind the chaining property is that the ability to do
$U$ and $V$ separately should be \emph{at least} as powerful as the
ability to do $UV$.  This property is always satisfied by $K_\Delta$,
and while it may not always be true for the metric-based strength
measures, it turns out to be true for a large class of them.

The second property, stability, just says that if we imagine adding an
extra qubit to our system and doing nothing to it, that should not
change the strength of the operation.  That is,
\begin{eqnarray}
K(U \otimes I) = K(U).
\end{eqnarray}
For example, if $U$ is a two-qubit unitary operator, then this type of
stability implies that the three-qubit strength of $U \otimes I$ is
the same as the two-qubit strength of $U$.  Provided an appropriate
measure of $n$-party entanglement is chosen, $K_\Delta$ can be shown
to satisfy this property.  The metric-based measures do not always
satisfy stability, but they do in some instances.

The third and final property, locality, just states that a strength
measure should be zero for products of local unitary operations.  That is,
\begin{eqnarray}
K(A \otimes B \otimes \ldots) = 0.
\end{eqnarray}
This is true of both $K_\Delta$ and the metric-based measures.

Well, imagine that we have such a strength measure, and we want to
perform a particular unitary operation, $U$, using a circuit
containing just controlled-{\sc not} and single-qubit unitary gates.
Imagine the circuit contains $M$ controlled-{\sc not}s.  Then,
applying our three properties, we see that the strength of $U$ can be
no more than the sum of the strengths of all the individual
controlled-{\sc not}s, plus the strengths of the local unitaries,
which are all zero.  This gives us an upper bound for the strength of
$U$, namely $M$ times the strength of the controlled-{\sc not},
\begin{eqnarray}
K(U) \leq M K(\mbox{{\sc cnot}}).
\end{eqnarray}
This, in turn, tells us that the number of controlled-{\sc not}s needed
in the circuit was at least the strength of $U$, divided by the
strength of the controlled-{\sc not},
\begin{eqnarray}
M \geq \frac{K(U)}{K(\mbox{{\sc cnot}})}.
\end{eqnarray}
Because of the stability property the strength of the controlled-{\sc
  not} is a constant, so we see that if the strength of $U$ scales
superpolynomially, then so must the number of gates needed to do $U$.

The reason this is an interesting line of thought is because nobody
has ever succeeded in proving nontrivial lower bounds on the
complexity of problems like the travelling-salesman problem.  Thus one
reason for being interested in measures of dynamic strength is that by
developing good measures of dynamic strength it might become possible
to prove some interesting lower bounds on computational complexity.

There are many other interesting questions and potential applications
of this idea of quantifying the dynamic strength of a quantum
operation, far too many for me to even describe the questions here,
much less the answers.  An attempt at fleshing out the theory of
dynamic strength may be found in~\cite{Nielsen02a}.  The key point,
however, is that by introducing measures of strength for quantum
dynamical operations we may be able to obtain insight into the
enormously complicated space of dynamical processes.

Let me conclude by looking again at the big picture.  My belief is
that the major scientific task of quantum information science is to
develop tools that will lead to insight into the properties of complex
quantum systems.  In quantum mechanics we're like chess players who've
just learnt the rules of the game, and are still trying to figure out
all the emergent properties those rules imply~\cite{Nielsen00g}.
We're doing so by developing overarching theories, like the theory of
entanglement and of dynamic strength, which let us understand ever
more complex phenomena.  I expect that as these theories are further
developed they will enable us to better understand complex systems,
not only in information processing, but also in other areas of
many-body physics.

\section*{Acknowledgements}

The point of view described in this paper owes a lot to many enjoyable
conversations with Jennifer Dodd and Tobias Osborne.  Thanks also to
Andrew Childs and Tobias Osborne for their feedback on the manuscript.


\end{document}